\documentclass[english,two column]{revtex4-1}
\usepackage[T1]{fontenc}
\usepackage[latin9]{inputenc}
\usepackage{bm}
\usepackage{color}
\usepackage{amsmath}
\usepackage{amssymb}
\usepackage{graphicx}
\usepackage{babel}
\begin{document}
\title{Dynamical excitations of one-dimensional Fulde-Ferrell pairing Fermi superfluid}

\author{Peng Zou$^1$}
\author{Huaisong Zhao$^1$}
\email {hszhao@qdu.edu.cn}
\author{Feng Yuan$^1$}
\author{Shi-Guo Peng$^2$}
\email {pengshiguo@wipm.ac.cn}

\affiliation{$^1$College of Physics, Qingdao University, Qingdao 266071, China}
\affiliation{$^2$State Key Laboratory of Magnetic Resonance and Atomic and Molecular
Physics, Innovation Academy for Precision Measurement Science and
Technology, Chinese Academy of Sciences, Wuhan 430071, China}

\begin{abstract}
We theoretically investigate a one-dimensional Fulde-Ferrell Fermi
superfluid at a finite effective Zeeman field $h$, and study entire dynamical excitations related to density perturbation. By
calculating the density dynamic structure factor, we find anisotropic
dynamical excitations in both collective modes and single-particle
excitations. Along the direction of centre-of-mass momentum $p$,
there are two obvious gapless collective modes with different
speed. The lower collective modes is from the usual gauge symmetry
breaking and has a larger speed than the one in the negative direction
of $p$. The higher one is due to the direction spontaneous symmetry breaking of centre-of-mass momentum $p$,
and separates two kinds of single-particle excitations in the positive
$p$ direction. However, this higher mode disappears in the opposite direction of $p$, where two single-particle excitations overlap
with each other. These signals of dynamical excitations can do help
to distinguish Fulde-Ferrell superfluid from the conventional Bardeen-Cooper-Schrieffer
superfluid in the future experiment.
\end{abstract}
\maketitle

\section{Introduction}

Since the discovery of superconducting phenomena in mercury by K.
Onnes in 1911, it is realized that particles with opposite momentum
and spin can generate a molecular Cooper pair carrying zero center-of-mass
(COM) momentum to decrease energy. This conventional Bardeen-Cooper-Schrieffer
(BCS) superfluidity originates from the Bose-Einstein-Condensation
(BEC) of Cooper pairs at zero momentum. Later it is theoretically
predicted by Fulde and Ferrel \citep{FuldePR1964} and Larkin and
Ovchinnikov \citep{Larkin1964} that Cooper pairs can also carry a
finite COM momentum, and condense at a nonzero momentum $p$ with
order parameter in either a plane-wave FF-type $\Delta\left(\bm{r}\right)=\Delta e^{i\bm{p}\centerdot\bm{r}}$ or
a standing-wave LO-type $\Delta\left(\bm{r}\right)=\Delta{\rm cos}\left(\bm{p}\centerdot\bm{r}\right)$.
In such an ansatz, the two mismatched Fermi surfaces of different
spin components can overlap, thereby supporting a spatially inhomogeneous
superfluidity. These pairing states generate an exotic Fulde-Ferrell-Larkin-Ovchinnikov
(FFLO) superfluidity \citep{CasalbuoniRMP2004}. 

These FFLO superfluids have been actively searched for almost six
decades. In condensed matter physics, some strong signals of FFLO
states come from the research in heavy fermions \citep{RadovanNAT2003,GloosPRL1993,HuxleyCM1993,BianchiPRL2003,MartinPRB2005}
and organic superconductors \citep{Croitoru2017CM}, which are only
indirect experimental evidences. Recently several experimental works
reported evidence of pair-density-wave in high-Tc superconductor
by scanning tunnelling microscopy technique \citep{LiuNAT2023,ZhaoNAT2023,AishwaryaNAT2023},
which ignites the passion again and provides new possibilities to search and study
FFLO state. An ultracold atomic Fermi gas has proven to be an ideal
tabletop system for the pursuit of FFLO superfluidity \citep{RadzihovskyRPP2010}.
The FFLO state is thought to be very fragile in three dimensions and
has quite narrow parameter space. Both lower spatial dimension \citep{LiuPRA2007}
and spin-orbit coupling effect \citep{HuNJP2013,LiuPRA2013} in Fermi
superfluid have been reported that they can expand the parameter space and theoretically
investigate the possible phase diagrams related to FFLO state. Also
a theoretical strategy via a Dark-State Control of Feshbach resonance
was proposed to realize FF superfluid in ultracold atomic gases \citep{HePRL2018}. 

In fact it is expected that the full dynamical excitations of a certain
matter state can be utilized to check the existence of this state, or distinguish it from other states. Dynamic
structure factor is an important many-body physical quantity, and includes rich information related to
dynamical excitations of the system \citep{Bloch2008rmp,Giorgini2008rmp}.
Experimentally dynamic structure factor can be directly measured
by a two-photon Bragg scattering technique, which had been used to
investigate dynamical excitations of the BCS-BEC crossover
Fermi superfluid, including the single-particle excitations \citep{VeeravalliPRL2008},
collective Goldstone phonon mode \citep{HoinkNP2017}, second sound
\citep{LiSCI2022} and Higgs mode \citep{DykeArxiv2023}. A spin-charge
separation of repulsive one-dimensional (1D) Fermi gas has also been
studied by this technique \citep{SenaratneSCI2022}. Thus it is interesting
to study the dynamical excitations of an FFLO Fermi superfluid, and
find its dynamical characters related to its symmetry structure.

In this paper, we theoretically investigate a 1D spin-polarized Fermi
superfluid in an FF-type pairing state, and discuss its entire dynamical excitations by numerically calculating
the dynamic structure factor of this system with random phase
approximation \citep{AndersonPR1958,LiuPRA2004}. We will initially
discuss the state of equation of the system in FF superfluid,
and then introduce all possible dynamical excitations, which may provide some background knowledge to anisotropic Josephson effect related to FFLO state.  This paper
is organized as follows. In the next section, we use the motion equation
of Green\textquoteright s function to introduce the microscopic model
of a 1D spin-polarized interacting Fermi gas, and outline the mean-field approximation, and then show how to calculate the response function with random phase approximation. We give results of the dynamic structure factor of
FF superfluids in Sec. III, and we give our conclusions and outlook
in Sec. IV. Some calculation details are listed in the Appendix.

\section{Methods}

\subsection{Model and Hamiltonian}

We consider a uniform 1D spin-polarized Fermi gases
with s-wave contact interaction. The system can be described by a
model Hamiltonian 
\begin{equation}
H=\sum_{k\sigma}\left(\epsilon_{k}-\mu_{\sigma}\right)c_{k\sigma}^{\dagger}c_{k\sigma}+g_{1D}\sum_{pkk'}c_{k\uparrow}^{\dagger}c_{p-k\downarrow}^{\dagger}c_{p-k'\downarrow}c_{k'\uparrow},
\end{equation}
here $\epsilon_{k}=k^{2}/2m$ is the dispersion relation of spin-$\sigma$
free particles with mass $m$ in reference to the chemical potential
$\mu_{\sigma}$, and $c_{k\sigma}(c_{k\sigma}^{\dagger})$ is their
annihilation (generation) operator in momentum representation. Here
and in the following we have set physical quantities $\hbar=k_{B}=1$
for convenience. The interaction $g_{1D}=-\gamma n_{0}/m$ describes an attractive interplay between opposite spin components with a dimensionless
quantity $\gamma$. Since we consider a uniform system with bulk density
$n_{0}$, the inverse of Fermi wave vector $k_{F}=\pi n_{0}$/2 and
Fermi energy $E_{F}=k_{F}^{2}/2m$ are used as length and energy units, respectively.
Usually the difference of chemical potential is used to define an
effective Zeeman field $h=\left(\mu_{\uparrow}-\mu_{\downarrow}\right)/2$,
and $\mu=\left(\mu_{\uparrow}+\mu_{\downarrow}\right)/2$ is the average
chemical potential.

\begin{figure}
\includegraphics[scale=0.35]{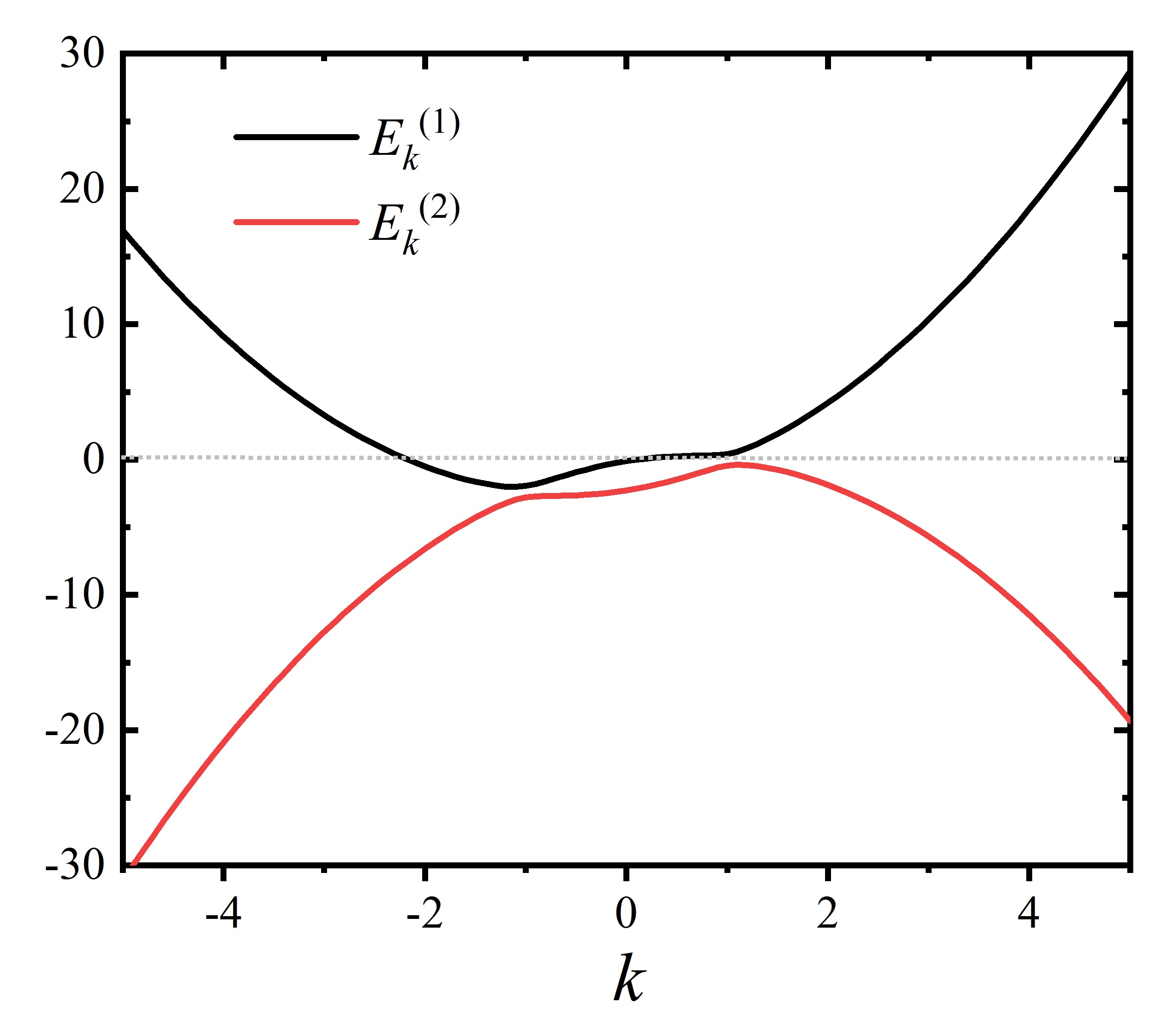}\caption{Two branches of single-particle excitation spectrum at interaction strength $\gamma=3$
and effective Zeeman field $h=1.2E_{F}$. Here COM momentum $p=1.18k_F$.}
\label{fig:spectrum}
\end{figure}

At zero temperature $T=0$, usually the system will comes into a
superfluid state, in which two opposite-spin atoms form a molecular Cooper
pair to decrease energy. Interestingly, an FF-type superfluid, who has a non-zero COM momentum in
Cooper pair, possibly turns out and even becomes the ground state
at a non-zero effective Zeeman field $h$. In a standard mean-field
treatment, the order parameter of FF superfluid can be expressed
in a plane-wave form $\Delta\left(x\right)\equiv\Delta e^{ipx}$,
where COM momentum $p$ and pairing gap $\Delta=g_{1D}\sum_{_{k}}\left\langle c_{-k+p/2\downarrow}c_{k+p/2\uparrow}\right\rangle $
are two important degrees of freedom of order parameter $\Delta\left(x\right)$. Within
this approximation, a mean-field Hamiltonian is expressed as 

\begin{equation}
\begin{array}{cl}
H_{\rm{mf}} & = \sum_{k}\left(\xi_{k}-h\sigma_{z}\right)c_{k\sigma}^{\dagger}c_{k\sigma}-\Delta^{2}/g_{1D}\\
 & -\sum_{k} \Delta \left(c_{-k+p/2\downarrow}c_{k+p/2\uparrow}+h.c.\right),
\end{array}
\end{equation}
where $\xi_{k}=\epsilon_{k}-\mu$. The exact solution of mean-field
Hamiltonian $H_{{\rm mf}}$ can be obtained by motion equation of
Green's functions. Finally we get three Green's functions, whose expressions
are listed below

\begin{equation}
G_{1}\equiv\left\langle \left\langle c_{k+p/2\uparrow}|c_{k+p/2\uparrow}^{\dagger}\right\rangle \right\rangle =\underset{l}{\sum}\frac{\left[G_{1}\right]_{k}^{l}}{\omega-E_{k}^{l}},\label{eq:G1}
\end{equation}

\begin{equation}
G_{2R}\equiv\left\langle \left\langle c_{-k+p/2\downarrow}^{\dagger}|c_{-k+p/2\downarrow}\right\rangle \right\rangle =\underset{l}{\sum}\frac{\left[G_{2R}\right]_{k}^{l}}{\omega-E_{k}^{l}},\label{eq:G2R}
\end{equation}

\begin{equation}
\varGamma\equiv\left\langle \left\langle c_{k+p/2\uparrow}|c_{-k+p/2\downarrow}\right\rangle \right\rangle =\underset{l}{\sum}\frac{\left[\varGamma\right]_{k}^{l}}{\omega-E_{k}^{l}},\label{eq:Gam}
\end{equation}
The double brackets in above equations are used to define Fourier transformation of double-time Green's function. Here it should be emphasized that the expression of spin-down Green's functions $G_{2R}$ is different from the one of spin-up Green's function $G_1$. Their definitions closely depend on their way coupled to pairing Green's function $\Gamma$ when
solving motion equation of Green's function. All expressions related
to$\left[G_{1}\right]_{k}^{l}$, $\left[G_{2R}\right]_{k}^{l}$ and
$\left[\Gamma\right]_{k}^{l}$ will be listed in the last appendix.
$l=1,2$ denotes two branches of quasi-particle energy spectrum $E_{k}^{(1)}$
and $E_{k}^{(2)}$, 

\begin{equation}
E_{k}^{(1,2)}=\frac{kp}{2m}-h\pm E_{k},\label{eq:spectrum}
\end{equation}
where $E_{k}=\sqrt{\xi_{kp}^{2}+\Delta^{2}}$ and $\xi_{kp}=\epsilon_{k}+\epsilon_{p}/4-\mu$. In typical parameters used in this paper,
we find that the value of $E_{k}^{(2)}$ is always negative, while
$E_{k}^{\left(1\right)}$ can be either positive or negative.
The distributions
of these two spectra are shown in Fig.~\ref{fig:spectrum}, in which $E_{k}^{\left(1\right)}$
is negative when $-2.17k_{F}<k<0.12k_{F}$. 

The mean-field thermodynamic potential of the system reads 
\begin{equation}
\begin{array}{cl}
\Omega = & -  \Delta^{2}/g_{1D}+\sum_{k}\left(\xi_{k}-E_{k}\right) \\
 &+ \sum_{k}T\left[\rm{ln} \it{f}\left(-E_{k}^{(\rm{1})}\right)+\rm{ln} \it{f}\left(E_{k}^{(\rm{2})}\right)\right],
\end{array}\label{eq:TDpotential}
\end{equation}
which is connected to the free energy by $F=\Omega+\mu N$. Here $f\left(x\right)=1/\left(e^{x/T}+1\right)$
is the Fermi-Dirac distribution function at temperature $T$. In this paper, we focus our discussion on an almost zero temperature ($T=0.01T_F$), to avoid an unnecessary numerical divergence induced by zeros of $E_k^{(1)}$.  

\subsection{State of equations}

The values for average chemical potential $\mu$, amplitude of order
parameter $\Delta$ and COM momentum $p$
can be respectively determined with minimization of thermodynamic potential $\Omega$ in Eq.~\ref{eq:TDpotential} to them, namely $N=-\partial\Omega/\partial\mu$, $\partial\Omega/\partial\Delta=0$
and $\partial\Omega/\partial p=0$. These three relations respectively give the total particle number equation
\begin{equation}
N=\sum_{k}\left(1-\frac{\xi_{kp}}{E_{k}}\right)+\sum_{k}\frac{\xi_{kp}}{E_{k}}\left[f\left(E_{k}^{(1)}\right)+f\left(-E_{k}^{(2)}\right)\right],\label{eq:density}
\end{equation}
pairing gap equation 
\begin{equation}
\frac{\Delta}{g_{1D}}=\sum_{k}\frac{\Delta}{2E_{k}}\left[f\left(E_{k}^{(1)}\right)+f\left(-E_{k}^{(2)}\right)-1\right],\label{eq:gap}
\end{equation}
and COM momentum equation 

\begin{widetext}
\begin{equation}
\sum_{k}\left[p\left(1-\frac{\xi_{kp}}{E_{k}}\right)+\left(2k+\frac{\xi_{kp}}{E_{k}}p\right)f\left(E_{k}^{(1)}\right)-\left(2k-\frac{\xi_{kp}}{E_{k}}p\right)f\left(-E_{k}^{(2)}\right)\right] 
=0. \label{eq:momentum}
\end{equation} 
\end{widetext}
The value of $\mu$, $\Delta$ and $p$ should be self-consistently solved with Eqs.~\ref{eq:density}, \ref{eq:gap} and \ref{eq:momentum}.

\begin{figure}
\includegraphics[scale=0.33]{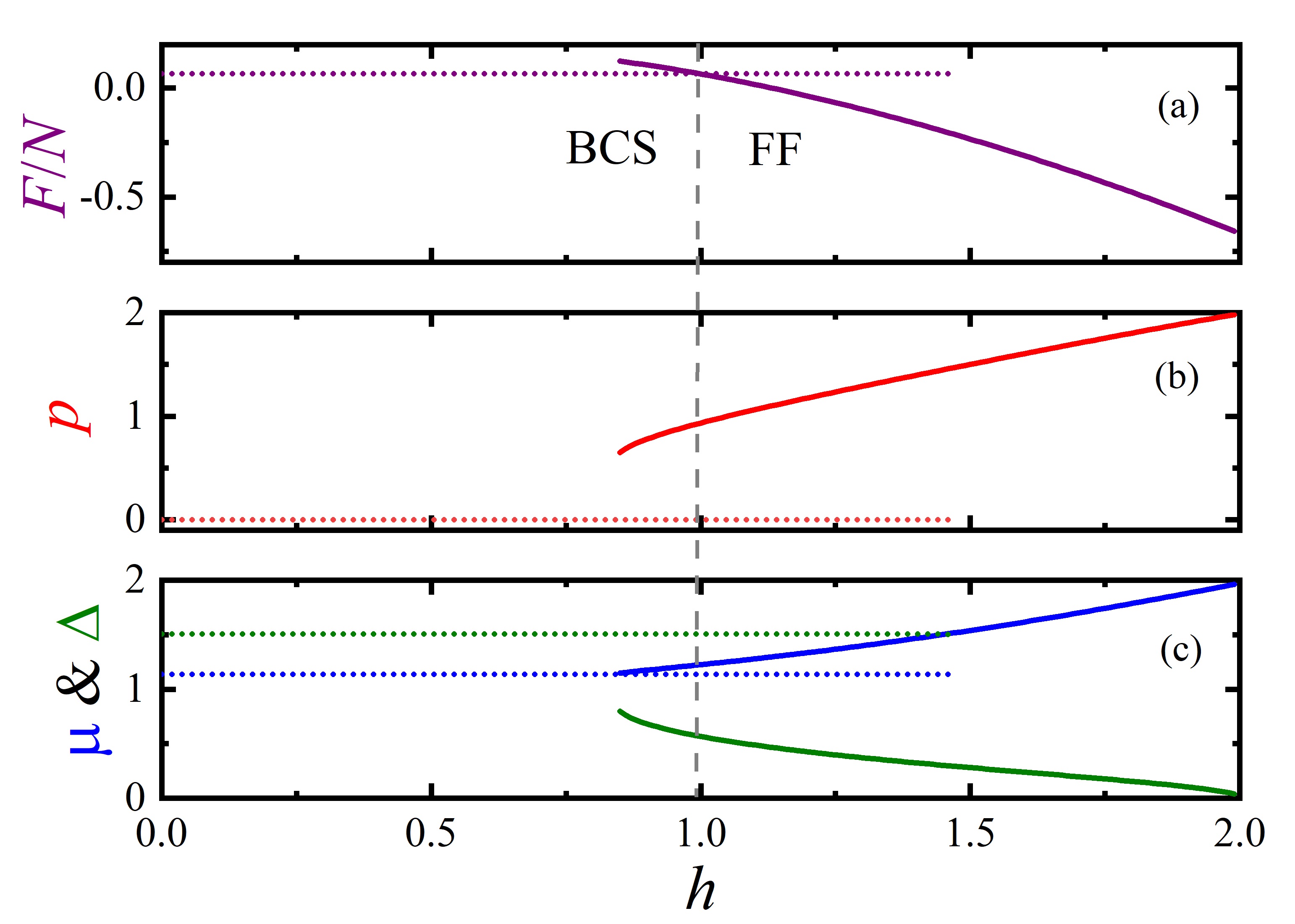}\caption{State of equation at interaction strength $\gamma=3$. Three panels show the curves of (a) free energy, (b)
COM momentum and (c) average chemical potential and order parameter
at different effective Zeeman field, respectively. }
\label{fig:soe3}
\end{figure}

\begin{figure}
\includegraphics[scale=0.35]{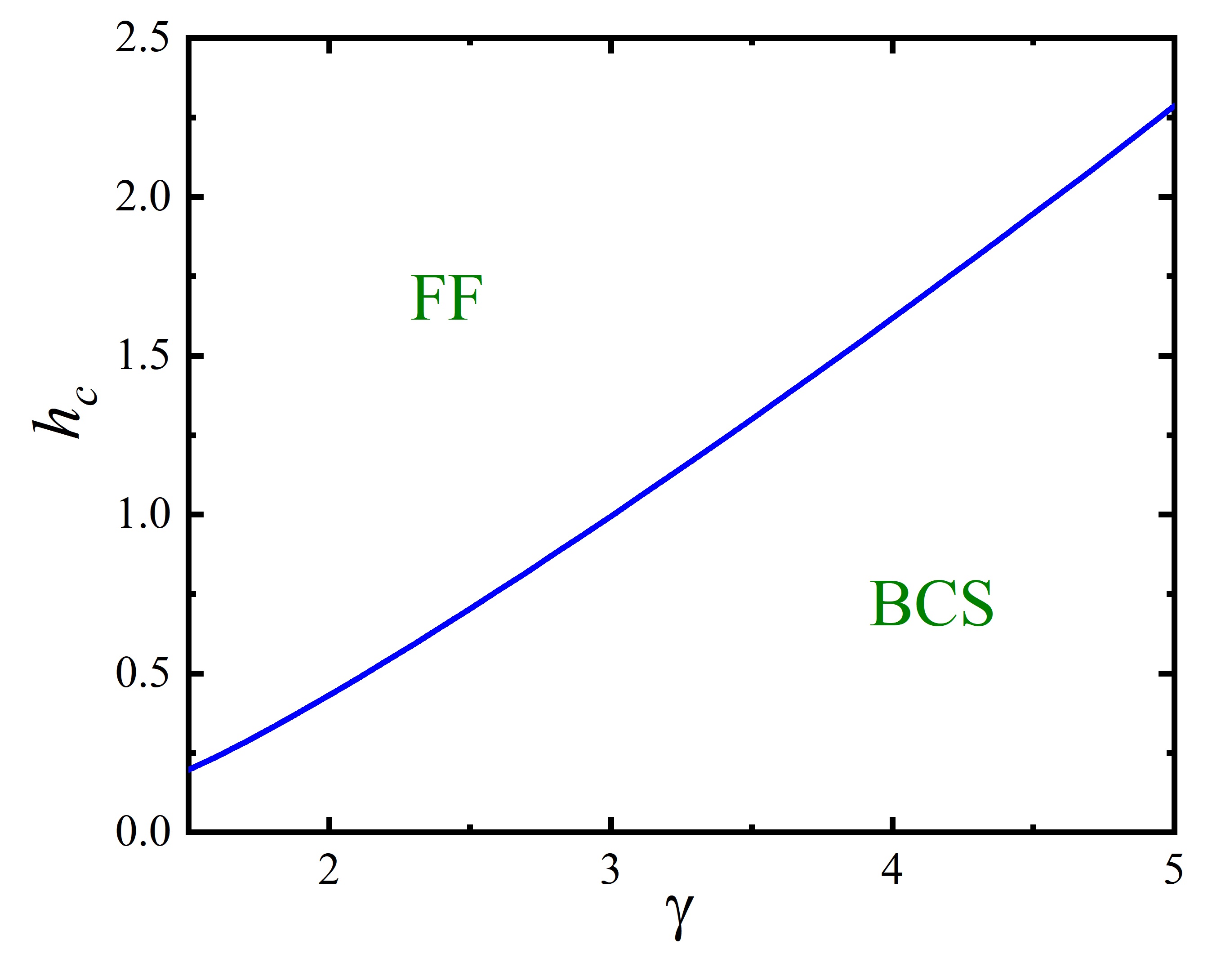}\caption{The critical effective Zeeman field $h_{c}$ at different interaction
strength $\gamma$.}
\label{fig:hc}
\end{figure}

The possible stable state of a system is determined by all minima in free energy $F$. Generally there are three possible states here. Besides the trivial normal state, who always has a zero pairing gap $\Delta$ and the largest free energy, the conventional BCS superfluid and FF superfluid are the other two states with relatively lower free energy.    
As shown in Fig.~\ref{fig:soe3}, at a small $h$, the conventional
BCS superfluid is the ground state of the system, whose COM momentum
$p=0$. The value of $h$ can hardly influence the chemical potential
$\mu$ and pairing gap $\Delta$, which can be thought as the analogy of the Meissner
effect in the superfluid. Interestingly an FF superfluid with
a non-zero COM momentum $p$ begins to turn out when $h$ is large enough, whose free energy is larger than the one of BCS superfluid. When $h$ comes to a critical value $h_{c}$ (a little smaller than $1.0E_{F}$ here),  this FF superfluid shares the same free energy as BCS superfluid. Further increasing $h$, FF superfluid will replace
BCS superfluid to be the new ground state of the system for $h>h_{c}$.
This phase transition had been introduced in Ref. \citep{LiuPRA2007}. Here $\Delta$
and $p$ are two necessary degree of freedom of in a FF-type order
parameter $\Delta\left(x\right)=\Delta e^{ipx}$. Pairing gap $\Delta$ is related to the conventional gauge symmetry breaking between normal state and superfluidity, and the direction of COM momentum $p$   is related to another spontaneous symmetry breaking since $\pm p$ corresponding the same free energy. While $\Delta$
always decreasing with $h$, $p$ shows a monotonically increasing behaviour
with $h$. The different dependence behavior of $\Delta$ and $p$ with effective Zeeman field $h$ goes on reflecting that they come from different symmetry breaking and may induce different collective modes.  It is easy to know that the phase transition shown in Fig.~\ref{fig:soe3} is a first-order one, which is manifested by the discontinuous behavior of $\mu$, $\Delta$ and $p$. We have checked that the
same phase transition also happens at different interaction strength
$\gamma$ with different critical Zeeman field $h_{c}$, whose value is shown in Fig.~\ref{fig:hc}.
A larger interaction strength $\gamma$ requires a bigger critical effective Zeeman field $h_c$ to make the system come into the FF superfluid state. 

\subsection{Calculation of dynamical excitations}

When an interacting system comes into a superfluid state, usually
there will be four typical densities. Besides the normal spin-up
density $n_{1}=\left\langle \psi_{\uparrow}^{\dagger}\psi_{\uparrow}\right\rangle $
and spin-down density $n_{2}=\left\langle \psi_{\downarrow}^{\dagger}\psi_{\downarrow}\right\rangle $,
the pairing physics of opposite-spin atoms generates the other anomalous
density $n_{3}=\left\langle \psi_{\downarrow}\psi_{\uparrow}\right\rangle $
and its conjugate counterpart $n_{4}=\left\langle \psi_{\uparrow}^{\dagger}\psi_{\downarrow}^{\dagger}\right\rangle $.
These pairs form phase coherent Cooper pairs with zero or finite
COM momentum $p$. The interaction between particles makes
these four densities couple closely with each other. Any fluctuation in each kind of density will influence other densities and generate an obvious density fluctuation of them. Also any weak perturbation potential $V_{\textrm{pert}}$ will generate density fluctuations $\delta n$, and they are connected with each other by response function $\chi$, namely $\delta n=\chi V_{\textrm{pert}}$, in the frame of linear response theory.

The entire dynamical excitations of the system consist of possible collective
excitations and single-particle excitations, which are also mainly
connected to the physical properties of Cooper pairs. The phase spontaneously
breaking of them generates a gapless Goldstone collective mode, while
the breaking of Cooper pairs forming parts of single-particle excitation.
These dynamical excitations can be well described by the density dynamic
structure factor, which is from the imaginary part of related response
functions $\chi$. The direct calculation of dynamic structure factor suffers
from the many-body difficulty. A potential approximation to overcome
this problem is the random phase approximation, which is firstly brought
by P. W. Anderson and has be verified to be a qualitatively reliable way
to study dynamical excitations. In our previous work, we had introduced
how to calculate dynamic structure factor with random phase approximation
\citep{GaoPRA2023,ZhaoPRA2023}. This approximation finds the connection
between the beyond mean-field response function $\chi$ and its mean-field
approximation $\chi^{0}$, whose calculation is relatively easier,
by the following equation

\begin{equation}
\chi=\frac{\chi^{0}}{1-\chi^{0}M_{I}g_{1D}},\label{eq:RPA}
\end{equation}
where the constant matrix $M_{I}=\sigma_{0}\otimes\sigma_{x}$ is
the direct product of unit matrix $\sigma_{0}$ and Pauli matrix $\sigma_{x}$,
reflecting the coupling situation of four types of densities. $\chi^{0}$
is a four dimension matrix in mean-field level, and its $ij$ matrix
element $\chi_{ij}^{0}$ reflects the interaction-induced coupling
between density $n_{i}$ and $n_{j}$. When the system comes into
the FF superfluid, $\chi^{0}$ has nine independent matrix elements,
namely
\begin{equation}
\chi^{0}=\left[\begin{array}{cccc}
\chi_{11}^{0} & \chi_{12}^{0} & \chi_{13}^{0} & \chi_{14}^{0}\\
\chi_{12}^{0} & \chi_{22}^{0} & \chi_{23}^{0} & \chi_{24}^{0}\\
\chi_{14}^{0} & \chi_{24}^{0} & -\chi_{12}^{0} & \chi_{34}^{0}\\
\chi_{13}^{0} & \chi_{23}^{0} & \chi_{43}^{0} & -\chi_{12}^{0}
\end{array}\right].\label{eq:kai0}
\end{equation}
After Fourier transformation to above response function, we obtain
the expression of all matrix elements in the momentum-energy representation.

With Eqs.~\ref{eq:RPA} and \ref{eq:kai0}, we can obtain the expression
of total-density response function  $\chi_{n}\equiv\chi_{11}+\chi_{22}+\chi_{12}+\chi_{21}$.
Based on the fluctuation and dissipation theorem, its imaginal part
is  connected to density dynamical structure
factor by 

\begin{equation}
S_{n}=-\frac{1}{\pi}\frac{1}{1-e^{-\omega/T}}{\rm Im}\left(\chi_{n}\right).
\end{equation}

\section{Results}

\begin{figure}
\includegraphics[scale=0.3]{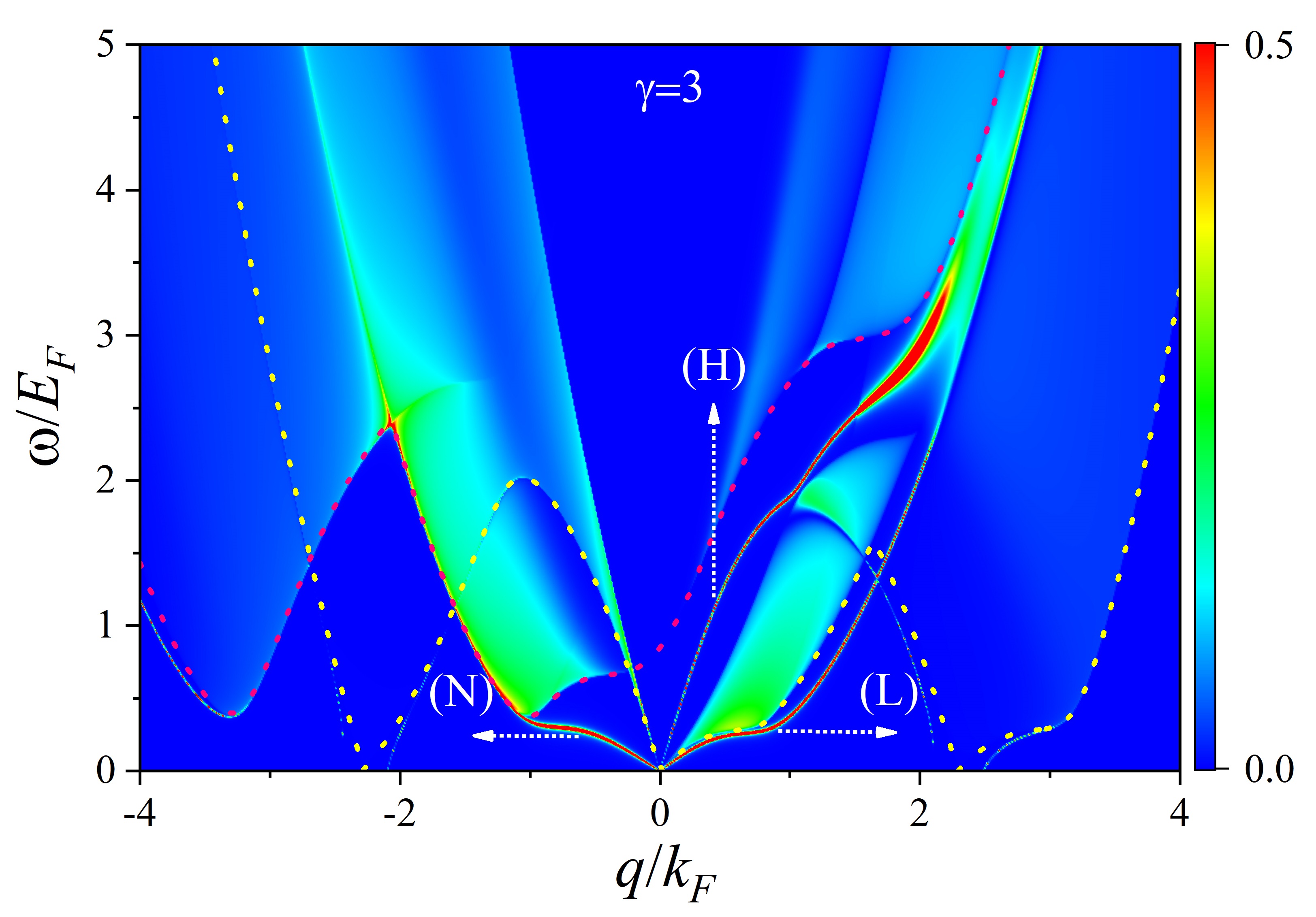}\caption{The density dynamic structure factor $S_{n}(q,\omega)$ of FF superfluid at interacting strength $\gamma=3$. (L) the lower collective phonon mode along the COM momentum $p$ direction, (H) the higher collective phonon mode, (N) the collective phonon mode in the opposite direction of $p$. }
\label{fig:dsf_color}
\end{figure}

We numerically calculate the density dynamic structure factor $S_{n}(q,\omega)$
of an FF superfluid at interaction strength $\gamma=3$. We define the direction of COM momentum $p$ is along the positive direction of transferred momentum $q$. The results
are shown in Fig.~\ref{fig:dsf_color}. Intuitively we see an anisotropic
dynamical behavior. The dynamical excitation at a positive transferred
momentum $q$ is different from the one at a negative $q$. This is
due to the direction dependence of the non-zero COM momentum $p$
in FF superfluid. Along the direction of COM momentum $p$ (namely
$q>0$), we see two kinds of gapless collective mode
(curves marked by (L) and (H)), and two-separated regimes of single-particle
excitation. When $q<0$, the higher gapless collective mode
disappears, and two kinds of single-particle excitations overlap with
each other. The speed of gapless collective phonon mode (curve marked
by (N)) is a little smaller than that in the positive direction. In
the following, we will separately introduce collection modes
and single-particle excitations.

\subsection{Collective excitation}

The origin of collective mode is closely related to the symmetry breaking
of a certain matter state. A gapless collective excitation comes from
a certain spontaneous symmetry breaking of the system. It is interesting
to notice that there are two gapless collective modes
at a positive transferred momentum $q$, which are also displayed
at Fig.~\ref{fig:dsf_color}. The lower mode is the conventional collective
Goldstone phonon mode (marked by (L)), which requires the lowest excitation
energy among all possible excitations. Its physical origin is due to
the gauge symmetry breaking of the phase of order parameter $\Delta(x)$.
The higher gapless collective mode (marked by (H)) is due
to the symmetry breaking of the direction of COM momentum $p$, which is continuous symmetry breaking in higher dimension system but not continuous in 1D system due to its specific spatial dimension. A similar mode as the higher collective mode here is also reported
in a LO-type superconductor with order parameter 
$\Delta\left(x\right)=\Delta\rm{cos}\left(\it{px}\right)$, which is called the gapless
Higgs mode since its amplitude of order parameter displays spatial
periodic variation. However, this should be different from the one in FF superfluid, whose amplitude is always a constant value to keep continuous translational symmetry. So the higher collective mode in FF superfluid we argue that it is better to be called a gapless phonon-like mode, instead of a gapless Higgs mode.

\begin{figure}
\includegraphics[scale=0.35]{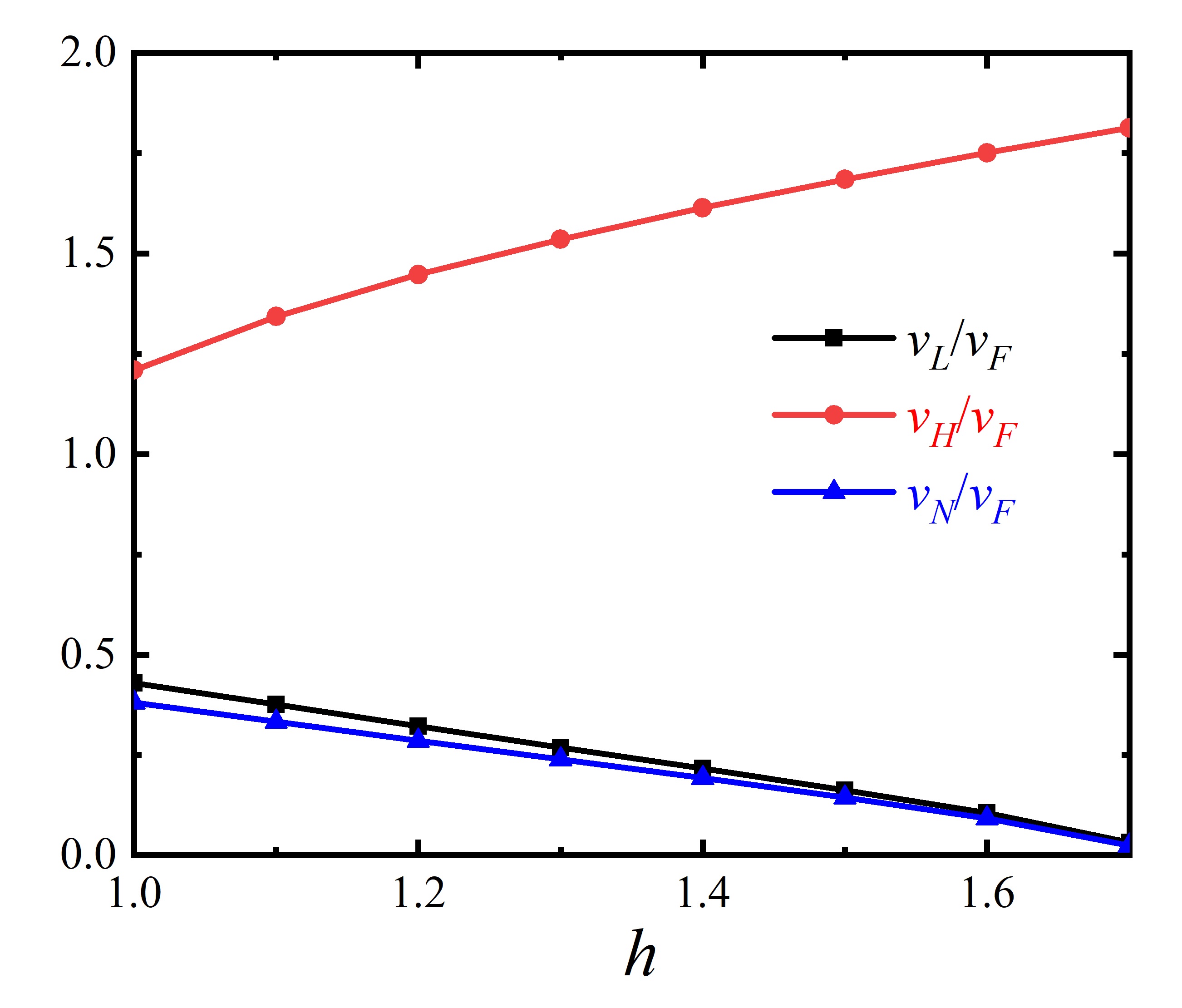}\caption{Speeds of three collective modes at different effective Zeeman field $h$. $v_L$ and $v_H$ are respectively the speed of lower and higher branch collective mode along the direction of COM momentum $p$, while $v_N$ is the speed of collective mode in the opposite direction of $p$.}
\label{fig:speed}
\end{figure}

The other difference from the BCS superfluid is that the lowest collective phonon mode displays an anisotropic excitation between the positive
and negative direction, which is due to the direction dependence of COM momentum $p$ in FF superfluid.
The similar anisotropic dynamics of phonon mode is also reported
in a spin-polarized Fermi system in square optical lattice \citep{HeikkinenPRA2011}.
From Fig.~\ref{fig:speed}, it is easy to know that the velocity of
phonon mode in a negative $p$ direction (marked by (N)) is always smaller
than the one in the positive $q$ direction. Also the absolute value
of phonon velocity decreases with $h$ in the negative $q$ direction. 

 The effective Zeeman field $h$-dependence of the speed for three collective modes is shown in Fig.~\ref{fig:speed}, which displays that both $v_{L}$ and $v_{N}$ decrease with $h$ and slowly goes to zero at a large enough $h$. Finally $\Delta$ also 
touch zero and the superfluid disappears when $h$ is around $2.0E_F$. However, $v_{H}$ always rises with $h$, and this behavior is consistent with that of COM
momentum $p$ (panel (b) in Fig.~\ref{fig:soe3}). The similar $h$-dependence between $v_H$ and $p$ again demonstrates a close connection between them. Also the different $h$-dependence between $v_L$ and $v_N$  demonstrates that the two gapless collective modes come from different symmetry breaking mechanism, and are related to different fluctuation of order parameter.

\subsection{Single-particle excitation}

Now we discuss the pair-breaking excitation, which is an important
part of single-particle excitation and takes up large regimes
in Fig.~\ref{fig:dsf_color}. This excitation is usually a continuous
excitation and its lowest excitation energy is determined by
the single-particle spectrum (Eq.~\ref{eq:spectrum}). To understand
all possible ways of pair-breaking excitation, it is better to understand
this physics by part of expression in response function $\chi^{0}$,
namely

\begin{equation}
\frac{f\left(E_{p}^{l}\right)-f\left(E_{p+q}^{l'}\right)}{i\omega_{n}+E_{p}^{l}-E_{p+q}^{l'}}.\label{eq:SE-possible}
\end{equation}
Here $E_{p}^{l}$ should consider all possible combinations
of single-particle spectrum. As shown in Fig.~\ref{fig:spectrum}, $E_{k}^{(2)}$ is always negative, and $E_{k}^{\left(1\right)}$ is positive except in a narrow regime $-2.17k_{F}<k<0.12k_{F}$. The possible pair-breaking
excitation may happen requiring that the numerator of Eq.~\ref{eq:SE-possible}
can not be zero, and also $E_{p}^{l}-E_{p+q}^{l'}<0$ to ensure a
positive excitation energy. 

Finally we find two possible pair-breaking excitations,
namely $\left|E_{k}^{(1)}-E_{k+q}^{(1)}\right|$ (11-type excitation)
at a limited regime of momentum $k$ and $\left|E_{k}^{(2)}-E_{k+q}^{(1)}\right|$(21-type
excitation) at a full regime of $k$. And their minima is the lowest
energy to break an FF-type Cooper pair, which are labeled by both
pink and yellow dotted lines in Fig.~\ref{fig:dsf_color}, respectively.
It should emphasized that the 11-type excitations (dotted yellow line) is a gapless pair-breaking
excitation and is absent in the conventional BCS superfluid. Honestly there is slightly bias between predictions from yellow dotted lines and random phase approximation's prediction when $q$ is around $\pm 2.5k_F$, which may due to the limited excitation regime of momentum $k$ in 11-excitation.
The minimum of 21-type excitation is labeled with a dotted pink line.
For a positive transferred momentum $q$, these two pair-breaking
excitations are separated with each other by just the higher gapless
collective mode. However, they are mixed with each other
in a negative transferred momentum $q$. These differences can do help
to find the higher collective mode in future experiment,
and distinguish FF-type superfluid from BCS superfluid.

\section{Conclusions and outlook}

In summary, we theoretically calculate the dynamic structure factor
of a 1D FF superfluid with random phase approximation to study dynamical excitations of the system. We find an anisotropic dynamical
excitation between positive and negative directions in both collective
modes and single-particle excitations. In the positive direction, we
find two gapless collective modes. The lower one comes from
the spontaneous breaking of gauge symmetry, while the higher one comes
from the direction symmetry breaking of COM momentum $p$. The sound speed in the positive
direction is larger than the one in the negative direction. There
are two types of pair-breaking excitations, and one of them is a gapless
excitation, which is absent in the BCS superfluid. In the positive
direction, these two kinds of pair-breaking excitations are just separated by the higher gapless collective mode, but overlap with each other
in the negative direction. These dynamical excitations can do help
to distinguish FF superfluid from the conventional BCS superfluid in future experiment. 

\section{Acknowledgements}
We are grateful for discussions with Wei Yi, Xiaoquan Yu and Hui Hu. This research was supported by National Natural Science Foundation of China under Grants No. 11804177 (P.Z.), No. 11547034 (H.Z.), Grant No. 11974384 (S.-G.P.) and 12374250 (S.-G.P.), and the National Key R\&D Program under Grant No. 2022YFA1404102 (S.-G.P.).

\section{Appendix}
In this appendix, we list expressions of three Green's functions and mean-field response function $\chi^{0}$. The first Green's functions is $G_{1}\left(k,\omega\right)=\sum_{l}\left[G_{1}\right]_{k}^{l}/\left(\omega-E_{k}^{l}\right)$, 
with 
\begin{equation}
\begin{array}{cc}
\left[G_{1}\right]_{k}^{(1)}=\frac{E_{k}^{(1)}+\xi_{k-p/2}+h}{E_{k}^{(1)}-E_{k}^{(2)}}, & \left[G_{1}\right]_{k}^{(2)}=-\frac{E_{k}^{(2)}+\xi_{k-p/2}+h}{E_{k}^{(1)}-E_{k}^{(2)}}.\nonumber\end{array}
\end{equation}
The second one is $G_{2R}\left(k,\omega\right)=\sum_{l}\left[G_{2R}\right]_{k}^{l}/\left(\omega-E_{k}^{l}\right),$
with
\begin{equation}
\begin{array}{cc}
\left[G_{2R}\right]_{k}^{(1)}=\frac{E_{k}^{(1)}-\xi_{k+p/2}+h}{E_{k}^{(1)}-E_{k}^{(2)}}, & \left[G_{2R}\right]_{k}^{(2)}=-\frac{E_{k}^{(2)}-\xi_{k+p/2}+h}{E_{k}^{(1)}-E_{k}^{(2)}}.\nonumber\end{array}
\end{equation}
The third Green's function is $\varGamma\left(k,\omega\right)=\sum_{l}\left[\varGamma\right]_{k}^{l}/\left(\omega-E_{k}^{l}\right),$
with
\begin{equation}
\begin{array}{cc}
\left[\Gamma\right]_{k}^{(1)}=-\frac{\Delta}{E_{k}^{(1)}-E_{k}^{(2)}}, & \left[\Gamma\right]_{k}^{(2)}=\frac{\Delta}{E_{k}^{(1)}-E_{k}^{(2)}}.\nonumber\end{array}
\end{equation}

The expressions of all nine independent matrix elements in mean-field
response function $\chi^{0}$ are respectively

$\chi_{11}^{0}=+\underset{pll'}{\sum}\left[G_{1}\right]_{p}^{l}\left[G_{1}\right]_{p+q}^{l'}\frac{f\left(E_{p}^{l}\right)-f\left(E_{p+q}^{l'}\right)}{i\omega_{n}+E_{p}^{l}-E_{p+q}^{l'}},$

$\chi_{12}^{0}=-\underset{pll'}{\sum}\left[\varGamma\right]_{p}^{l}\left[\varGamma\right]_{p+q}^{l'}\frac{f\left(E_{p}^{l}\right)-f\left(E_{p+q}^{l'}\right)}{i\omega_{n}+E_{p}^{l}-E_{p+q}^{l'}},$

$\chi_{13}^{0}=+\underset{pll'}{\sum}\left[G_{1}\right]_{p}^{l}\left[\varGamma\right]_{p+q}^{l'}\frac{f\left(E_{p}^{l}\right)-f\left(E_{p+q}^{l'}\right)}{i\omega_{n}+E_{p}^{l}-E_{p+q}^{l'}},$

$\chi_{14}^{0}=+\underset{pll'}{\sum}\left[\varGamma\right]_{p}^{l}\left[G_{1}\right]_{p+q}^{l'}\frac{f\left(E_{p}^{l}\right)-f\left(E_{p+q}^{l'}\right)}{i\omega_{n}+E_{p}^{l}-E_{p+q}^{l'}},$

$\chi_{22}^{0}=+\underset{kll'}{\sum}\left[G_{2R}\right]_{k}^{l}\left[G_{2R}\right]_{k+q}^{l'}\frac{f\left(E_{k}^{l}\right)-f\left(E_{k+q}^{l'}\right)}{i\omega_{n}+E_{k}^{l}-E_{k+q}^{l'}},$

$\chi_{23}^{0}=-\underset{kll'}{\sum}\left[\varGamma\right]_{k}^{l}\left[G_{2R}\right]_{k+q}^{l'}\frac{f\left(E_{k}^{l}\right)-f\left(E_{k+q}^{l'}\right)}{i\omega_{n}+E_{k}^{l}-E_{k+q}^{l'}},$

$\chi_{24}^{0}=-\underset{kll'}{\sum}\left[G_{2R}\right]_{k}^{l}\left[\varGamma\right]_{k+q}^{l'}\frac{f\left(E_{k}^{l}\right)-f\left(E_{k+q}^{l'}\right)}{i\omega_{n}+E_{k}^{l}-E_{k+q}^{l'}},$

$\chi_{34}^{0}=+\underset{kll'}{\sum}\left[G_{2R}\right]_{k}^{l}\left[G_{1}\right]_{k+q}^{l'}\frac{f\left(E_{k}^{l}\right)-f\left(E_{k+q}^{l'}\right)}{i\omega_{n}+E_{k}^{l}-E_{k+q}^{l'}},$

$\chi_{43}^{0}=+\underset{kll'}{\sum}\left[G_{1}\right]_{k}^{l}\left[G_{2R}\right]_{k+q}^{l'}\frac{f\left(E_{k}^{l}\right)-f\left(E_{k+q}^{l'}\right)}{i\omega_{n}+E_{k}^{l}-E_{k+q}^{l'}}.$

\end{document}